\documentclass[prl,aps,10pt,twocolumn]{revtex4}

\usepackage{latexsym}
\usepackage{amstext}
\usepackage{amsmath}
\usepackage[normalem]{ulem}
\usepackage{subfigure}
\usepackage{epsfig, graphicx}
\usepackage{color}
\usepackage{amsfonts,bm,color}
\usepackage{verbatim}
\usepackage{float}
\usepackage{latexsym}
\usepackage[svgnames]{xcolor}
\usepackage{colortbl}
\usepackage{float}
\usepackage{wrapfig}
\usepackage{multirow}
\newcommand{\beq}{\begin{eqnarray}}
\newcommand{\eeq}{\end{eqnarray}}
\newcommand{\beqq}{\begin{eqnarray*}}
\newcommand{\eeqq}{\end{eqnarray*}}
\newcommand{\p}{\partial}

\newcommand{\eps}{\varepsilon}

\newcommand{\x}{\mbox{\boldmath$x$}}

\newcommand{\y}{\mbox{\boldmath$y$}}

\newcommand{\w}{\mbox{\boldmath$w$}}

\newcommand{\n}{\mbox{\boldmath$n$}}

\newcommand{\ds}{\displaystyle}

\begin{document}
\title{Extreme escape from a cusp: when does geometry matter for the fastest Brownian particles moving in crowded cellular environments?}
\author{K. Basnayake$^1$, D. Holcman$^1$}
\affiliation{{$^1$ Group of data modeling and computational biology, Ecole Normale Sup\'erieure-PSL, Paris, France.}}
\date{\today}
\begin{abstract}
We study here the extreme statistics of Brownian particles escaping from a cusp funnel: the fastest Brownian particles among $n$ follow an ensemble of optimal trajectories located near the shortest path from the source to the target. For the time of such first arrivers, we derive an asymptotic formula that differs from the classical narrow escape and dire strait obtained for the mean first passage time. Consequently, when particles are initially distributed at a given distance from a cusp, the fastest do see some properties characterizing the cusp geometry. Therefore, when many particles diffuse around impermeable obstacles, the geometry plays a role in the time to reach a target. In the biological context of cellular transduction with signalling molecules, having to escape such cusp-like domains slows down fast signaling. To conclude, generating multiple copies of the same molecule helps bypass a crowded environment to transmit a molecular signal quickly.
\end{abstract}
\maketitle
\begin{figure}[http!]
	\center
	\includegraphics[scale=0.35]{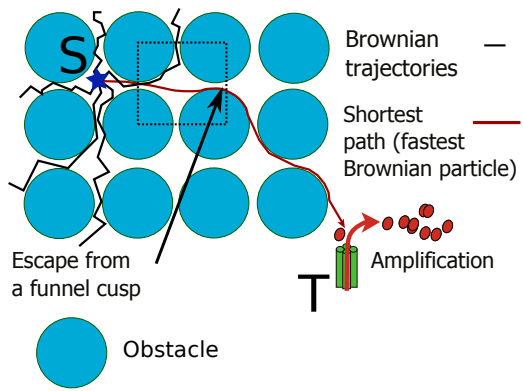}
	\caption{ \small {\bf Schematic representation of signaling activation using the fastest Brownian  messenger.} When the initial position of the source (S) of particles is distant from the cusp funnel of the obstacles, the trajectory of the fastest particle is an optimal path from the source to the target passing through several cusps. The many trajectories (black) have to avoid many impermeable obstacles (blue). At the target site (T), the first arriving particles activate a secondary pathway, leading to an amplification of the signal.}
\label{fig1}
\end{figure}

The redundancy principle describes the need to generate many copies of the same object to guarantee the successful execution of a biological function within a certain time constraint. Recently, it was shown that the time scale of biological signal transduction , such as the activation of a channel, a cascade of secondary messengers as well as gene activation by a transcription factor are mediated by first among many copies of particles (molecules, ions, transcription factors, etc.) that arrive at a small target \cite{schuss2019redundancy}. In this scenario, the classical Smoluchowski rate, which is extended in narrow escape theory \cite{schuss2007narrow,pillay2010asymptotic,cheviakov2010asymptotic,holcman2013control} only provides the statistics of the mean and becomes irrelevant for estimating the activation time.
The mean time taken by the fastest particles to find a small window (i.e. the binding site), has been computed for regular domains \cite{Holcman2015}.\\
In the probabilistic formulation, for $n$ i.i.d. Brownian particles in a bounded domain $\Omega$ with the boundary that is reflecting (except at a narrow window), the shortest arrival time is defined by $\tau^{(n)}=\min (t_1,\ldots,t_n),$
where $t_i$ are the arrival times of the $n$ paths in the ensemble. The first moments of $\langle\tau^{(n)}\rangle$ were computed in dimension one \cite{Redner,majumdar2007brownian,schehr2014exact,extreme1,extreme3,extreme5,extreme6,bray2013persistence}  and recently in higher dimensions \cite{basnayake2019asymptotic} leading to the following asymptotic formulas:
\beq \label{finalform}
({\bar{\tau}}^{(n)})^{dim 1} \approx& \frac{\delta^2}{4D\ln\left(\frac{n}{\sqrt{\pi}}\right)},
\eeq
\beq
({\bar{\tau}}^{(n)})^{dim 2}\approx& \ds \frac{\ds \delta^2}{\ds 4 D \log\left(\frac{\pi
\sqrt{2}n}{8\log\left(\frac{1}{a}\right)}\right)},
\eeq
Here $D$ is the diffusion coefficient, $\delta$ is the length of the shortest ray from the initial point to the small exiting window and $n$ is the number of particles that were initially released. The radius of the absorbing disk is $a$. Interestingly, the fastest particles use a trajectory close to the optimal path, showing the major contrast with the statistics of a typical brownian particle.\\
In the present study, we focus on the escape of $n$ Brownian particles from a two-dimensional domain with a funnel cusp, for various initial distributions. This is a ubiquitous modality of cellular signaling, where Brownian molecules should find their way across a crowded environment paved with many impermeable obstacles. In this case, the generic geometrical shape of a region between round obstacles located on a cellular membrane is a cusp funnel (Fig. \ref{fig1}). We recall that the formula for the mean time of a single Brownian particle to escape from such region, to the leading order \cite{holcman2011} is
\beq
\bar\tau=\frac{\pi|\,\Omega|}{ 2D\sqrt{\eps/R}}\left(1+O(\sqrt{\eps/R})\right)\hspace{0.5em}\mbox{for}\
\eps\ll1.
\eeq
The domain has a surface $|\Omega|$ while the cusp is with a size of $\epsilon$ in the opening and a curvature of $R$. The diffusion coefficient is $D$. The new asymptotic formula we derive here deviates from previous ones as the size of the narrow target appears outside the logarithmic term and the escape time for the fastest particle located inside the bulk is given by
\beq
\tau^{(n)} &\sim & \frac{16 \pi^2R^3}{16 \eps D \ds (\frac{1-\cos(c\sqrt{\tilde \eps})}{\tilde\eps})^2 \log(\frac{2n}{\sqrt{\pi}})}.
\eeq
Here $c$ is a constant that depends on the shape of the domain outside the cusp funnel and $\tilde \eps=\frac{\eps}{R}$. As N increases, the path chosen by the fastest approaches the shortest geodesic \cite{basnayake2018extreme}. This new formula has consequences for studying the time scales of molecular transduction occurring in crowded cellular membranes.  Moreover, it indicates that the notion of an ``effective'' diffusion coefficient is not useful to characterize the time scale of cellular activation that uses the fastest particles because the associated statistics notably differ from the mean properties of diffusion.\\

\begin{figure}[http!]
	\center
	\includegraphics[scale=0.22]{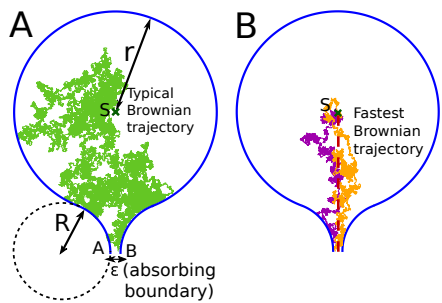}
\caption{{\small {\bf Escape path for a typical and the fastest of $n$ i.i.d. Brownian trajectories} {\bf Left.}. A Brownian trajectory starts initially at position S and escapes into a narrow cusp opening ($2\epsilon=|AB|$). $R$ is the radius of curvature. {\bf Right.} Examples of two trajectories for the fastest among $n=100$, starting at S.}}
\label{fig2}
\end{figure}

\begin{figure}[http!]
\center
\includegraphics[scale=0.16]{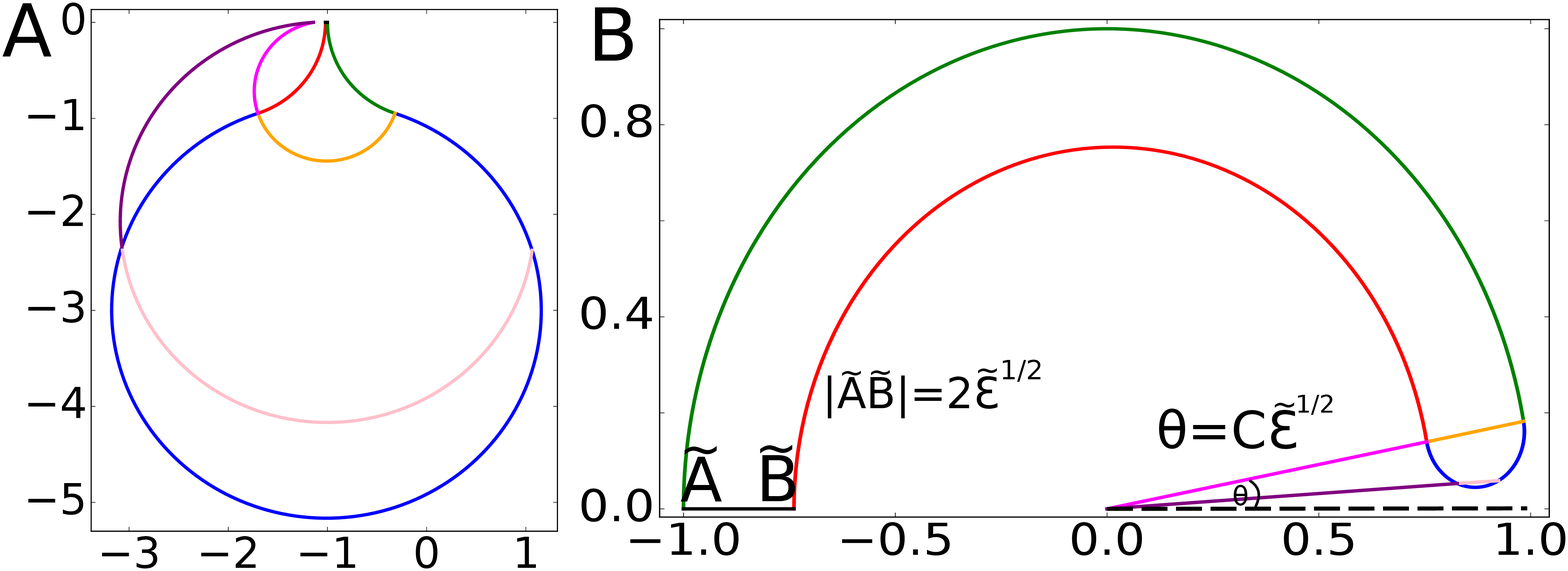}
\caption{{\small {\bf Mapping of the cusp-domain into a banana-shaped domain.}  {\bf Left.} Domain containing a cusp funnel. {\bf Right.} Mapped domain $w(\Omega)$. The yellow arc (left) is mapped in the end of the banana at an angle $c\sqrt{\tilde \eps}$, while the cusp is mapped into the red and green arcs.}}
\label{fig3}
\end{figure}

{\noindent \bf Computing the mean escape time from a funnel for the fastest Brownian particle.}
The distribution of exit times for the first Brownian particles to escape is
\beqq
\Pr\{\tau^{1}>t\} =  \left[ \Pr\{t_{1}>t\} \right]^N = \left[ \int_{\Omega} p(\x,t)\right]^N ,
\eeqq
and the mean is
\beq
\bar{\tau}^{(n)}= \int\limits_0 ^{\infty} \left[ \int_{\Omega} p(\x,t)\right]^N dt.
\eeq
The probability density function $p$ satisfies the Fokker-Planck equation (FPE):
\begin{align}\label{IBVP}
\frac{\p p(\x,t\,|\,\y)}{\p t} =&D \Delta p(\x,t\,|\,\y)\quad\hbox{\rm for } \x,\y \in \Omega,\\
p(\x,0\,|\,\y)=&\delta(\x-\y)\quad \hbox{\rm for } \x,\y \in\Omega\nonumber\\
\frac{\p p(\x,t\,|\,\y)}{\p \n} =&0\quad \hbox{\rm for } \x \in\p\Omega_r, \y\in\Omega\nonumber\\
p(\x,t\,|\,\y)=&0\quad \hbox{\rm for } \x \in \p\Omega_a,  \y\in\Omega,\nonumber
\end{align}
where $\Omega$ is a domain with a cusp funnel so that the boundary $\p\Omega$ contains a small absorbing window $\p\Omega_a$ of length $\eps$ at the cusp opening and the rest of the boundary is reflecting $\p\Omega_r=\p\Omega-\p\Omega_a$ (Fig. \ref{fig1}). We will focus here on the case where the initial condition is a function $p(\x,0\,|\,\y)=p_0(\x)$ concentrated inside the bulk at a certain distance from the cusp.\\
The domain $\Omega$ is generic for studying the effect of escape from narrow passages \cite{holcman2012brownian,holcman2011,holcman2013control} . To remove the cusp-singularity, we first normalize the domain by changing variable $z=x/R$, where $R$ is the curvature at the symmetric cusp (Fig. \ref{fig2}). The normalized radius is $\tilde \eps =\frac{\eps}{R}$. In the z-plan, we use the M\"obius conformal mapping
\begin{align}
w=w(z)=\frac{z-\alpha}{1-\alpha z},\label{w},
\end{align}
where $\alpha=-1-\sqrt{\tilde \eps}$, to map the cusp domain into the banana shape \cite{holcman2012brownian,Holcmanschuss2018} Fig. \ref{fig3}.
Setting $p(z,t)=v(w,t)$, and using the polar coordinates as $w=Re^{i\theta}$, the system (\ref{IBVP}) is converted to:
\beq
\frac{\p v(w,t)}{\p t}
&=& \frac{D|w(1-\sqrt{\tilde\eps})-1|^4}{(4\tilde\eps+O(\tilde\eps^{3/2}))R^2
} \frac{\p^2 v(w,t)}{\p \theta^2} \hspace{0.5em} \mbox{ for }\w\in\Omega_w\nonumber \\
\frac{\p v(w)}{\p n}&=&\,0\hspace{0.5em}\mbox{ for } w\in\p\Omega_w-\p\Omega_{w,a}\label{PDEw} \\
v(w)&=&\,0\hspace{0.5em}\mbox{for}\ w\in\p\Omega_{w,a}.\nonumber
\eeq
The asymptotic solution of equation (\ref{PDEw}) is independent of the radial variable $R$ to the leading order.
The reflecting boundary condition is given at the angle $\theta_r=c\sqrt{\tilde\eps}$,  where $c=O(1)$ is a constant independent of $\tilde\eps$ to the leading order.
The boundary conditions are \cite{holcman2012brownian}
\begin{align}
\frac{\p v(c\sqrt{\eps},t)}{\p \theta}=0,\quad v(\pi,t)=0.\label{BCs}
\end{align}
The diffusion equation, therefore, simplifies to:
\beq
\frac{\p v(\theta,t)}{\p t}= \eps a(\theta)  \frac{\p^2 v(\theta,t)}{\p \theta^2},
\eeq
where the diffusion tensor accounts for the cusp geometry, given by
{\small
\beqq
a(\theta)&=& \frac{D}{4 R^2} \left(\frac{|e^{i\theta}-1-e^{i\theta}\sqrt{\tilde \eps}|^4}{4{\tilde\eps}^2}\right)\\&=&\frac{D}{4R^2} (1-\sqrt{\tilde\eps}+O(\tilde\eps))^2(\frac{1-\cos(\theta)}{\tilde\eps})^2.
\eeqq}
The initial condition is such that $v(c\sqrt{\eps},t)$ tends to $\delta_{c\sqrt{\eps}}$, as $t$ tends to zero. We search for a WKB-type solution \cite{Schuss2} of the form:
\beq \label{wkb1}
v(\theta,t)= \frac{K_{\tilde\eps}(\theta,t)}{\sqrt{\tilde\eps}}\exp(-\frac{\Psi(\theta,t)}{\tilde\eps}).
\eeq
Here the function $\Psi$ satisfies the Eikonal equation
\beq
-\frac{\p \Psi}{\p t}=a(\theta)\left(\frac{\p\Psi}{\p\theta}\right)^2.
\eeq
The Taylor's expansion along the optimal trajectory gives
{\small
\beqq
\Psi(\tilde\theta(t),t,\theta_0)\approx \Psi(\tilde\theta(t),0,\theta_0)+\frac{1}{2}\Psi_{\theta \theta}(\tilde\theta(t),0,\theta_0) (\tilde\theta(t)-\theta)^2 +..,
\eeqq
}
where $\theta_0=c\sqrt{\tilde \eps}$. Due to the normalization condition of WKB-type solution, the zeroth order can be chosen $\Psi(\theta,0,\theta_0)=0$. Since $\Psi(\theta,t)$ is minimal along the optimal path $\tilde\theta(t)$, $\frac{\p \Psi}{\p x}(\tilde\theta(t),t)=0$. Thus the optimal solution satisfies:
\beq
\dot{{\tilde\theta}}&=&0\\
\tilde\theta(0)&=&c\sqrt{\eps},
\eeq
leading to $\tilde\theta(t)=c\sqrt{\eps}$.For computing the second-order term of the expansion, we note that $P(t,\theta_0)=\frac{2}{\ds \Psi_{\theta \theta}(\tilde\theta(t),0,\theta_0)}$ is a solution of
\beq
\frac{\p P(t,\theta_0)}{\p t} =2a(\tilde\theta(t,\theta_0)),
\eeq
leading to
\beq
P(t,\theta_0) &=&2\int_0^t a(\theta(u,\theta_0))du=2ta(c\sqrt{\tilde \eps})\\
&=&2t\frac{D}{4R^2} (1-\sqrt{\tilde\eps}+O(\tilde\eps))^2(\frac{1-\cos(c\sqrt{\tilde \eps})}{\tilde\eps})^2 .\nonumber
\eeq
For computing the remaining function $K_{\tilde\eps}$ in Eq. \ref{wkb1}, we note that $K_{\tilde\eps}(t,\theta_0)$ is a solution of the transport equation \cite{Schuss2}:
\beq
\frac{\p K_{\tilde\eps}(t,\theta_0)}{\p t}=-\left( \frac{a(\tilde\theta(t))}{P(t,\theta_0)}\right)K_{\tilde\eps}(\tilde\theta(t,\theta_0)).
\eeq
The solution is
\beq
K_{\tilde\eps}(\tilde\theta(t),\theta_0)=\frac{ K_{0}(\theta,c\sqrt{\tilde \eps})}{\sqrt{P(t,\theta_0)}}.
\eeq
Finally, using the normalization condition in the transformed domain, we obtain the outer-solution
(that does not satisfy the boundary conditions) to the leading order:
\beq
v(\theta,t,c\sqrt{\tilde \eps})= \frac{1}{\sqrt{4 \tilde \eps \pi\tilde D t} }\exp(-\frac{(\theta-c\sqrt{\tilde \eps})^2}{4\eps \tilde Dt} ),
\eeq
where $\tilde D=\frac{D}{4R^2} (1-\sqrt{\tilde\eps}+O(\tilde\eps))^2(\frac{1-\cos(c\sqrt{\tilde \eps})}{\tilde\eps})^2$.\\
We construct the entire solution $v_{ent}$ that accounts for the boundary condition, using the reflection principle in the interval of length $L=2l=2(\pi-\sqrt{\tilde \eps})$,
{\small
\beqq
{\scriptscriptstyle v_{ent}(\theta,t,\theta_0)=  \frac{K_0}{\sqrt{4 \tilde \eps \pi\tilde D t} } }
\sum_{n} \left( e^{(-\frac{(\theta-\theta_0+2Ln)^2}{4\eps \tilde Dt}} )-e^{(-\frac{(\theta+\theta_0+2Ln)^2}{4\eps \tilde Dt} )}
\right).
\eeqq
}
Here $K_0$ is a normalization constant computed using the survival probability $S(t)=\int_{\Omega} p(\x,t)d\x $ when the particles start inside the bulk, yielding:
{\small
\beq\label{survival}
S(t)=\frac{2}{R^2}\sqrt{\tilde \eps} \int_{\tilde\Omega} v_{ent}(\theta,t,c\tilde \eps) |w'(e^{i\theta})|^2 d\theta.
\eeq
}and $|w'(e^{i\theta})|^2 =\frac{2{\tilde \eps}}{(1-\cos(\theta))}$. We recall that the diffusion solution  $v_{ent}(\theta,t,c\tilde \eps)$ should be normalized such that $S(t) \to 1 $ as $t \to 0$. The Laplace's method applied to Eq.\ref{survival} for large values of $n$, when $\tilde \eps \ll 1$ shows that the weight $|w'(e^{i\theta})|^2$ does not matter in the leading order.\\
Using computations developed in \cite{basnayake2019asymptotic}, we obtain for an initial point at $\theta_0=c \sqrt{\tilde \eps}$, integrating from 0 to $\pi-c \sqrt{\tilde \eps}$
\beqq
 {\scriptstyle S(t)\approx \frac{K_0}{\sqrt{4 \tilde \eps \pi\tilde D t}  } \int \left( \frac{ e^{(-\frac{(\theta-\pi+c \sqrt{\tilde \eps})^2}{4\eps \tilde Dt})}-e^{(-\frac{(\theta+\pi-c \sqrt{\tilde \eps})^2}{4\eps \tilde Dt} )}}{(1+\cos(\theta+2c \sqrt{\tilde \eps))^2}} \right)d\theta }.
\eeqq
Thus $S(t)^n\approx \exp n\left(\ln(1-\frac{8\sqrt{\tilde \eps \tilde D t}}{L\sqrt{\pi}}e^{-L^2/(16(\tilde \eps \tilde D)t)})\right)$ and the mean escape time for the fastest among $n$ is
{\small\beq\label{formulaA}
\tau^{(n)}\approx \frac{(\pi-\sqrt{\tilde \eps})^2}{4 \tilde \eps \tilde D \log(\frac{2n}{\sqrt{\pi}})}
\approx  \frac{\pi^2R^3}{16 \eps D \ds (\frac{1-\cos(c\sqrt{\tilde \eps})}{\tilde\eps})^2 \log(\frac{2n}{\sqrt{\pi}})}.
\eeq
}
This result is obtained for an initial position where a Brownian particle starts inside the bulk \cite{basnayake2018extreme}.
\begin{figure}[http!]
\center
\includegraphics[scale=0.1]{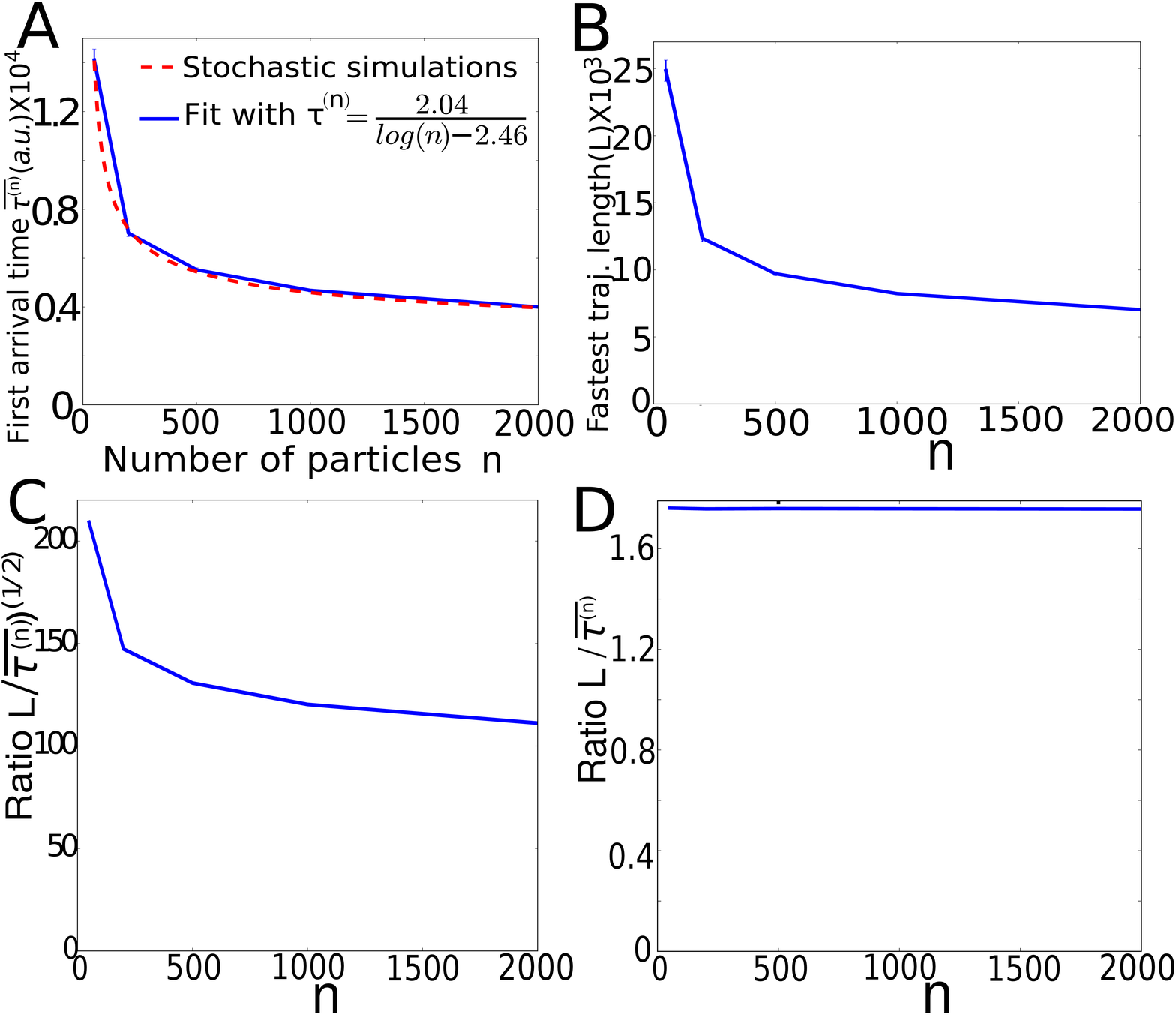}
\vspace{-1em}
\caption{ \small {\bf Arrival time statistics vs number of particles n (A)} First arrival times from stochastic simulations (blue) fitted with  eq. \ref{formulaA} (red). Parameters are $r=200$, $R=100$ and $\eps$=10 in the geometry of Fig. \ref{fig2} computed over 350 trials.
{\bf(B)} Approximated lengths of the trajectories. {\bf(C) and \bf(D)} Ratios $\frac{L_n}{\tau^{(n)}}$ and $\frac{L_n}{(\tau^{(n)})^2}$ vs $\eps$ and $n$.}
\vspace{-2em}
\label{figresult1}
\end{figure}
To test the predictions of this formula, we performed numerical simulations of 2D-Brownian motion of particles inside  a cusp funnel geometry (Fig. \ref{fig2}). The particles follow the Langevin's equation at the Smoluchowski's limit  (Brownian motion) $\dot{X}=\sqrt{2D}\dot{w}$, where $w$ is the Wiener noise. The discretized form is simulated using the Euler’s scheme: $X_n=X_{n-1}+\sqrt{2D \Delta t}\cdot\eta$, where $X_n$ is the position of the particle at the $n^\mathrm{th}$ time step while $\eta$ is a two-dimensional normal random variable. In all realisations, the diffusion coefficient $D$ and the width of the time step $\Delta t$ were chosen as one.\\
We find a good agreement of the log-dependency of $\tau^{(n)}$ with respect to the target size $\eps$ (Fig.\ref{figresult1}), predicted analytically in eq. \ref{formulaA}. Moreover, we notice that the ratio between the mean time $\tau^{(n)}$ and the trajectory length $L_n$ of the fastest particles is constant with respect to $n$, showing that trajectories behave as geodesics and not random walk paths, as the ratio $\frac{L_n}{(\tau^{(n)})^2}$ deviates from a constant (Fig. \ref{figresult1}C \& D).\\
We recall that during the simulation results (Fig.\ref{figresult1}), all particles were initially positioned at the center $S$ (Fig.\ref{fig2}). Next we varied the initial conditions (Fig.\ref{intialconditions}). One way to study this is to use 
the M\"obius transformation, that mapped the domain inner arcs into straight line segments in the banana-shaped transformed domain (Fig. \ref{fig3}) Orange \& Pink). Setting initial conditions on one of the arcs in the large circular domain of the bulk (Fig. \ref{intialconditions}A) is equivalent in the mapped domain to an initial positions located on a small straight segment.
\begin{figure*}[http!]
\center
\includegraphics[scale=0.17]{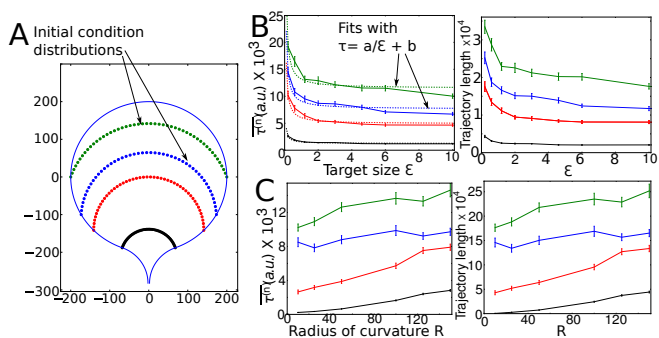} 
\caption{ \small {\bf First arrival times vs the approximated lengths of the fastest trajectories on the cusp geometry for various initial conditions.}  {\bf A.} 1000 particles were initially released simultaneously from each of the 50 points (dots) located on the arcs and simulated until the first one arrives to the target site.  {\bf B.} Time $\tau^{(n)}$ (Left) and approximated trajectory lengths of the fastest (Right) vs target size $\epsilon$ in the range 0.1-10.0. Individual fits are obtained with eq. \ref{formulaA}. {\bf C.} Time $\tau^{(n)}$ (Left) and approximated trajectory lengths $L_n$ of the fastest (Right) vs the cusp radius (R). Different initial conditions (colors) are used in correspondence between panel A, B and C. The error bars indicate $\pm$ standard error of the mean over 50 realisations. }
\vspace{-1em}
\label{intialconditions}
\end{figure*}
Using these different initial conditions, we confirm the agreement of the log-dependency of $\tau^{(n)}$ with respect to the size $\eps$ of the target (Fig.\ref{intialconditions}B Left: fits in dotted lines). Although we predicted that $\tau^{(n)}$ increases proportional to $R^3$ (Eq.\ref{formulaA}), numerical simulations show some deviations that depend on the initial position, that should be further explored. Moreover, we confirmed again in these simulations that the ratios between the mean time $\tau^{(n)}$ and the trajectory length $L_n$ of the fastest particles is constant with respect to both $\eps$ and $R$, while the ratio $\frac{L_n}{(\tau^{(n)})^2}$ is not constant, confirming that the trajectory of the fastest Brownian particle becomes deterministic as n increases.\\
{\noindent \bf Concluding remarks.}
We showed here that $n$ independent Brownian particles initially located away from a generic cusp funnel would escape following an ensemble of shortest paths. The parameters characterizing the cusp such as the curvature or the width of the window determine the mean time of the fastest, showing that the cusp geometry influences the escape time to leading order. When the initial locations of the particles are distributed far from the target, the dependency of $\tau^{(n)}$ with respect to the geometrical features remains an interesting question to further explore.\\
In the context of cell biology, the present modeling and results suggest that fast molecular signaling using the fastest Brownian particles between a source and target occurs along the shortest paths. The time to find the target depends on the shortest distance between obstacles creating funnel cusps especially on the surface membrane of living cells. 
This process of finding a target by the fastest defines the fast molecular activation of signal transduction, which is ubiquitous in cell biology: phototransduction, activation of G coupled receptors, and many more. The present results show that to estimate the activation time, it is not necessary to use an effective diffusion constant, but rather the properties of extreme statistics.
\bibliographystyle{apsrev4-1} 
\vspace{-3em}
\bibliography{bibliocusp}
\end{document}